\documentstyle[12pt]{article}
\newtheorem{theorem}{{\sc Theorem}}
\newcommand{\bt}{\begin{theorem}}
\newcommand{\et}{\end{theorem}}
\newcommand{\newsection}[1]{\setcounter{equation}{0} \setcounter{theorem}{0}
\section{#1}}

\newcommand{\NI}{\noindent}

\newcommand{\bea}{\begin{eqnarray}}
\newcommand{\eea}{\end{eqnarray}}

\def \b #1 {\bf #1}
\newcommand{\IR}{I\!\!R}
\newcommand{\IE}{I\!\!E}
\newcommand{\IC}{I\!\!C}

\newcommand{\IT}{I\!\!T}
\newcommand{\IZ}{Z\!\!\!Z}

\newcommand{\cla}{{\cal A}}

\newcommand{\cls}{{\cal S}}

\newcommand{\clf}{{\cal F}}
\newcommand{\clg}{{\cal G}}
\newcommand{\clh}{{\cal H}}
\newcommand{\clp}{{\cal P}}

\newcommand{\clb}{{\cal B}}

\newcommand{\cle}{{\cal E}}

\newcommand{\cln}{{\cal N}}
\newcommand{\cld}{{\cal D}}
\newcommand{\cll}{{\cal L}}

\newcommand{\clm}{{\cal M}}

\newcommand{\al}{\alpha}

\newcommand{\raro}{\rightarrow}

\newcommand{\sbs}{\subset}

\newcommand{\vsp}{\vskip 1em}

\newcommand{\ul}{\underline}

\def \qed {\hfill \vrule height6pt width 6pt depth 0pt}
\newcommand{\be}{\begin{equation}}
\newcommand{\ee}{\end{equation}}
\newcommand{\ben}{\begin{eqnarray*}}
\newcommand{\een}{\end{eqnarray*}}
\begin{document}
\thispagestyle {empty}
\sloppy

\centerline{\large \bf Markov shift in non-commutative probability }

\bigskip
\bigskip
\centerline{\bf Anilesh Mohari }
\smallskip
\centerline{\bf S.N.Bose Center for Basic Sciences, }
\centerline{\bf JD Block, Sector-3, Calcutta-91 }
\centerline{\bf E-mail:anilesh@boson.bose.res.in}
\bigskip
\smallskip
\centerline{\bf Abstract}
\smallskip
\bigskip
\vsp
We consider a class of quantum dissipative semigroup on a 
von-Neumann algebra which admits a normal invariant state. We investigate 
asymptotic behavior of the dissipative dynamics and their relation to that 
of the canonical Markov shift. In case the normal 
invariant state is also faithful, we also extend the notion of 
`quantum detailed balance' introduced by Frigerio-Gorini and prove that 
forward weak Markov process and backward weak Markov process are equivalent 
by an anti-unitary operator.    

\hfil\eject

\newsection{ Introduction:}

\vsp
Let $\tau=(\tau_t,\;t \ge 0)$ be a semigroup of identity preserving completely 
positive maps on a von-Neumann algebra $\cla_0$ acting on a separable 
Hilbert space $\clh_0$ and $\phi_0$ be an invariant normal state for 
$\tau$. We consider the unique minimal reversible system, 
constructed in [AcM2], i.e. a triplet $(\cla,\al_t,\phi)$,
where $\cla$ is a von-Neumann algebra acting on a Hilbert space $\clh$, 
$(\al_t,\;t \in \IT = \;\IR\; \mbox{or} \; \IZ )$ is a group of 
$*$-automorphism on $\cla$ and $\phi$ is an invariant state for 
$(\al_t)$, so that the following diagram

$$ (\cla,\phi) \;\;^{\al_t} \longrightarrow (\cla,\phi) $$
\be
j^f_0\;\uparrow	\hspace{2cm} \downarrow\;{\IE}_0
\ee
$$ (\cla_0,\phi_0) \;\;^{\tau_t} \longrightarrow (\cla_0,\phi_0)$$
commutes for all $t \ge 0$ where ${\it j^f_0}$ is an injective
$*$-homomorphism and $\IE_0$ is a completely positive map. Moreover
there exists a group of unitary operators $(S_t)$ on $\clh$ and a unit
vector $\Omega \in \clh$ so that $S_t\Omega=\Omega,\; \phi(X)=<\Omega,X \Omega>$and $\al_t(X)=S_t^*XS_t,\;\forall t \in \IR,\;X \in \cla$.   
Inspired by the classical notion $(S_t)$ will be referred as 
{\bf Markov shift }. 

\vsp
In case 
$(\cla_0,\tau_t,\phi_0)$ is itself a reversible system i.e. $\tau_t$ 
is also an endomorphism for each $t \ge 0$, 
then $\cla$ 
is isomorphic to $\cla_0$ and $(\alpha_t)$ is same as $(\tau_t)$ up to an
isomorphism. On the other hand, 
this dilation is indeed a generalization of Kolmogorov's 
construction of 
stationary Markov process in the non-commutative
frame-work [Da2,PaB,AcM2], where {\it weak Markov forward process } 
$(j^f_t$ $:\;t \in \IR)$ is defined by $j^f_t(x)=\alpha_t(j^f_0(x))$ $,\; 
\forall x \in \cla_0.$ The family of increasing 
projections $\{ j^f_t(I): \;t \in \IT \}$ is the non-commutative counterpart 
of the filtration generated by the process $(j^f_t:\cla_0 \raro \cla,\; t 
\in \IT)$. Furthermore in case $\IT=\IR$ and the map $t \raro \tau_t(x)$ is 
weak$^*$ continuous, then $(S_t)$ is also strongly continuous.  

\vsp
In this exposition we analyze asymptotic behavior of the weak Markov process 
$(j^f_t)$. Since the minimal reversible process is uniquely determined by 
the dynamical semigroup, it is expected that asymptotic behavior of $(j^f_t)$
will be related to that of $(\tau_t)$. At this point we remark very few general
results are known which guarantees existence of a normal invariant state. For a 
discussion and results on this issue we refer to [Da2,FaR1,FaR2]. For this 
exposition we assume existence of a normal invariant state and explore 
how ergodicity, mixing ( weak and strong ) of $(S_t)$ is related with that 
of $(\tau_t)$. We say the forward process is having {\bf Kolmogorov shift } 
or {\bf K-shift } property if the tail 
subspace is trivial, 
i.e. $j^f_t(I) \raro |\Omega><\Omega|$ strongly as $t \raro -\infty$.  
In particular we prove that the process is having Kolmogorov property 
if and only if $\phi_0(\tau_t(x)\tau_t(y)) \raro \phi_0(x)\phi(y)$ as 
$t \raro \infty$.
This notion was introduced in [AcM2] and explored its relation with 
the canonical commutation relation.  

\vsp
We investigate further the asymptotic behavior of the dynamical semigroup 
$(\tau_t)$ and to that end we assume $\phi_0$ to be also faithful. In a recent 
paper $[FaR3]$ Fagnola and Rebolledo found a useful criteria which guarantees 
faithful property of an invariant normal state in-terms of non existence of 
a non-trivial sub-harmonic projections. 
   
\vsp
We revisit Frigerio's original  work [Fr1] and introduce von-Neumann sub-algebras
${\cal F}= \{ x \in \cla_0:\;\tau_t(x^*)\tau_t(x)=\tau_t(x^*x),\;\tau_t(x^*)\tau_t(x)=\tau_t(xx^*),\; t \ge 0 \}$ and ${\cal I} =\{x \in \cla_0:\;\tau_t(x)=x,\; t \ge 0 \}$. It is obvious ${\cal I} \subset {\cal F}$. We prove that the equality 
${\cal F}={\cal I}$ is a sufficient condition for weak$^*$ limit 
$\tau_t(x) \raro E(x)$ as $t \raro \infty$ for any $x \in \cla_0$, where 
$E$ is the norm on projection on the von-Neumann sub-algebra 
$\{ x: \tau_t(x)=x,\;t \ge 0 \}$. 
This is a little 
improvement of Frigerio's work [Fr1] and in particular it removes
the asymmetric feature of Frigerio's original condition for strong mixing.
In this direction we added one important result which says how to get 
steady state which need not be faithful. In this regard we find the notion
of sub-harmonic projection introduced in [FR3] plays an important role.
 
\vsp
It is simple to note that any measure preserving strongly mixing 
flow does not satisfy this 
condition, thus this sufficient condition is not a necessary one 
for the shift $(S_t)$ to be strong mixing. 
Since $\{x \in \cla_0:\tau_t(x^*)\tau_t(x)=\tau_t(x^*x):\;t \ge 0 \}=\IC$
is a necessary condition for Kolmogorov's  property, 
in case strong mixing
is equivalent to Kolmogorov's property, Frigerio's criteria is also necessary 
for strong mixing. 
In this exposition we will show such equivalence if $\cla_0$ is a type-I 
von-Neumann algebra
with center completely atomic. In particular
strong mixing and K-shift property are equivalent if $\cla=\clb(\clh)$ or
$\cla=l^{\infty}(\cls)$, where $\cls$ is a countable set.

\vsp
In section 4, we explore further the faithful property of the invariant state
and 
consider the backward weak Markov process $j_t^b$ as in [AcM] associated with
a canonical adjoint quantum dynamical semigroup $(\tilde{\tau}_t)$. We also
consider the  
associated time reverse 
process $(\tilde{\cla}, \tilde{j}^f_t,\tilde{j}^b_t,\; \tilde{F}_{t]},\tilde{F}_{[t}, \tilde{
\Omega})$. There exists [AcM2] an anti-unitary operator $U_0:\clh \raro \tilde{\clh}$ which intertwines
the forward weak Markov process associated with $(\tau_t)$ to the backward 
weak Markov process associated with $(\tilde{\tau}_t)$. In particular we
check that ergodicity, weak mixing and strong mixing properties are time
reversible. However Kolmogorov's property seems to be delicate in the non-commutative
case.  

\vsp 
We also find that  
$\{x \in \cla_0: \tilde{\tau}_t\tau_t(x)=x,\;\forall t \ge 0 \}=
\{x \in \cla_0: \tau_t(x)=x,\;\forall t \ge 0 \}$ is 
a sufficient condition for weak$^*$ limit of $\tau_t(x) \raro E(x),\; 
\forall x \in \cla_0$ as $t 
\raro \infty$. Same is true if we interchange the role of $(\tau_t)$ with 
that of $(\tilde{\tau}_t)$. This condition seems to be weaker then that of 
Frigerio's modified condition.  
In case modular automorphism group associated with $\phi_0$ commutes with the 
dynamics $(\tau_t)$ this sufficient condition
is identical to that of the modified Frigerio's condition. 
We find this condition to be useful with the 
following implications:

\vsp
\NI (A) We prove that strong mixing and K-shift properties are equivalent when
$\cla_0$ is a type-I von-Neumann algebra with center completely atomic.
In such a case the following are equivalent:

\NI (1) $\{x \in \cla_0: \tau_t(x^*)\tau_t(x)=\tau_t(x^*x),\;\tau_t(x)\tau_t(x^*)=
\tau_t(xx^*),\; \forall t \ge 0 \}= \IC$, 

\NI (2) $\{x \in \cla_0: \tilde{\tau}_t\tau_t(x)=x,\; \forall t \ge 0 \} = \IC$,

\NI (3) $\phi_0(\tau_t(x)\tau_t(y)) \raro \phi_0(x)\phi_0(y)$ as $t \raro  
\infty$ for all $x,y \in \cla_0$. 

\vsp  
\NI Thus improved Frigerio's sufficient condition is also
necessary in this case for strong
mixing, equivalently for Kolmogorov's property.
 
\vsp
\NI (B) Inspired by seminal work [FrG] we also introduce a notion of 
`quantum detailed balance' 
and prove such an ergodic process is not only strongly mixing 
but also satisfies Kolmogorov's
property. Thus once more we found a quantum counter part of a 
well known classical result which says ergodicity and detailed balance 
give rise to a mixing system. Moreover there exists an anti-unitary operator
$R_0$ so that $R_0j_t^fR_0^*=j_{-t}^b,\;\forall \; t \in \IT$ and $R_0j^b_tR_0^*=\tilde{j}^f_{-t},\;\forall t \in \IT .$ Several model in quantum optics 
satisfies this detailed balance condition. However there are many 
interesting situation [Ma,MZ1,MZ2,MZ3] which suggests that the detailed 
balance condition for a Markov semigroup on quantum spin chain is far from
being understood and thus needs a better understanding, where the Hamiltonian 
dynamics do not commute with the dissipative dynamics. 

\vsp
We end this exposition with a short introduction to quantum mechanical
master equation and some implication of our results.

I wish to thank referee for his comments which not only make me aware of 
the related works but also helped me to revise the results and even
include new results.

\vsp
\newsection{ Stationary weak Markov process and shift: }

\bigskip
A family $(\tau_t,\;t \ge 0)$ of one parameter completely positive maps 
on $\cla_0$ with the properties $\tau_0=I,\;\tau_s \circ \tau_t=
\tau_{s+t},\;s,t \ge 0$ is called a {\it quantum dynamical semigroup}. 
If $\tau_t(I)=I,\;t \ge 0$ it is called a {\it Markov } semigroup. We say
a state $\phi_0$ is {\it invariant } for $(\tau_t)$ if $\phi_0(\tau_t(x))
= \phi_0(x)\;\forall t \ge 0.$

\vsp
Let $(\clh_0,\cla_0,\tau_t,t \ge 0,\phi_0)$ be a Markov semigroup 
and $\phi_0$ be an $(\tau_t)$-invariant state 
on $\cla_0$. We aim to recall from [AcM2] the quadruple $(\clh,\cla,\al_t,
\phi)$, where $\clh$ is a Hilbert space, $\cla$ is a von- Neumann 
algebra acting on $\clh$, $(\alpha_t,\;t \in \IR)$ is a group of 
automorphism on $\cla$ and $\phi$ is a normal state so that 
the diagram (1.1) commutes. The construction
goes along the line of Kolmogorov's construction of stationary
Markov processes or Markov shift with a modification [BhP] which 
takes care of the fact that $\cla_0$ need not be a commutative 
algebra. Here we review the construction given in [AcM2] in
order to fix the notations and important properties.

\vsp
We consider the class $\clm$ of $\cla_0$ valued functions
$\ul{x}: \IR \raro \cla_0$ so that $x_r \neq I$ for finitely
many points and equip with the point-wise multiplication
$(\ul{x}\ul{y})_r=x_ry_r$. We define the map $L: (\clm,\clm)
\raro \IC $ by
\be
L(\ul{x},\ul{y}) =
\phi_0(x_{r_n}^*\tau_{r_{n-1}-r_n}(x_{r_{n-1}}^*(.....x_{r_2}^*
\tau_{r_1-r_2}(x_{r_1}^*y_{r_1})y_{r_2})...y_{r_{n-1}})y_{r_n})
\ee
where $\ul{r}=(r_1,r_2,..r_n)\;r_1 \le r_2 \le .. \le r_n$ is
the collection of points in $\IR$ when either $\ul{x}$ or
$\ul{y}$ are not equal to $I$. That this kernel is well defined
follows from our hypothesis that $\tau_t(I)=I, \; t \ge 0$ and
the invariance of the state $\phi_0$ for $(\tau_t).$ The
complete positiveness of $(\tau_t)$ implies that the map $L$ is a
non-negative definite form on $\clm$. Thus there exists a
Hilbert space $\clh$ and a map $\lambda: \clm \raro \clh$ such
that $$<\lambda(\ul{x}),\lambda(\ul{y}) >= L(\ul{x},\ul{y}).$$
Often we will omit the symbol $\lambda$ to simplify our
notations unless more then one such maps are involved.

\vsp
We use the symbol $\Omega$ for the unique element in $\clh$
associated with $x=(x_r=I,\;r \in \IR )$ and the associated
vector state $\phi$ on $B(\clh)$ defined by $\phi(X)=<\Omega,X
\Omega>$.

\vsp
For each $t \in \IR$ we define shift operator $S_t: \clh \raro
\clh$ 
by the following prescription:
\be
(S_t\ul{x})_r = x_{r+t}
\ee
It is simple to note that $S = (( S_t ,\;t \in \IR))$ is a unitary
group of operators on $\clh$ with $\Omega$ as an invariant element.

\vsp
For any $t \in \IR$ we set
$$\clm_{t]}= \{\ul{x} \in \clm,\; x_r=I\;\forall r > t \} $$
and $F_{t]}$ for the projection onto $\clh_{t]}$, the closed
linear span of $\{\lambda(\clm_{t]})\}$. For any $x \in \cla_0$
and $t \in \IR$ we also set elements $i_t(x),\in \clm$ defined
by $$i_t(x)_r= \left \{ \begin{array}{ll} x ,&\; \mbox{if}\; r=t
\\ I,&\; \mbox{otherwise}\;  \end{array} \right.$$ So the map
$V_+: \clh_0 \raro \clh$ defined by $$V_+x=i_0(x)$$ is an
isometry of the GNS space $\{ x:<x,y>_{\phi_0}=\phi_0(x^*y) \}$
into $\clh$ and a simple computation shows that
$<y,V^*_+S_tV_+x>_{\phi_0}=<y,\tau_t(x)>_{\phi_0}$.  Hence
$$P^0_t=V^*_+S_tV_+,\;t \ge 0$$ 
where $P^0_tx=\tau_t(x)$ is a contractive semigroup of operators on the GNS
space associated with $\phi_0$. 

We also note that $i_t(x) \in \clm_{t]}$ and set $\star$-homomorphisms
$j^0_0: \cla_0 \raro \clb(\clh_{0]})$ defined by
$$
j^0_0(x)\ul{y}= i_0(x)\ul{y}
$$
for all $\ul{y} \in \clm_{0]}.$ That it is well defined follows
from (2.1) once we verify that it preserves the inner product whenever
$x$ is an isometry. For any arbitrary element we extend by linearity.
Now we define $j^f_0: \cla \raro \clb(\clh)$
by
\be
j^f_0(x)=j_0^0(x)F_{0]}.
\ee
Thus $j^f_0(x)$ is a realization of $\cla_0$ at time $t=0$ with
$j^f_0(I)=F_{0]}$. Now we use the shift $(S_t)$ to obtain the
process $j^f=(j^f_t: \cla_0 \raro \clb(\clh),\;t \in \IR )$ and
forward filtration $F=(F_{t]},\;t \in \IR)$ defined by the
following prescription:
\be
j^f_t(x)=S_tj^f_0(x)S^*_t\;\;\;F_{t]}=S_tF_{0]}S^*_t,\;\;t \in \IR.
\ee

\vsp
So it follows by our construction that $ j^f_{r_1}(y_1)j^f_{r_2}(y_2)...
j^f_{r_n}(y_n) \Omega = \ul{y}$ where $y_r=y_{r_i},\;$ if $r=r_i$
otherwise $I,\;(r_1 \le r_2 \le .. \le r_n)$.  Thus $\Omega$ is
a cyclic vector for the von-Neumann algebra $\cla$ generated by
$\{ j^f_r(x), \;r \in \IR, x \in \cla_0 \}$. From (2.4) we also
conclude that $S_tXS^*_t \in \cla$ whenever $X \in \cla$ and
thus we can set a family of automorphism $(\al_t)$ on $\cla$
defined by $$\al_t(X)=S_tXS^*_t$$ Since $\Omega$ is an invariant
element for $(S_t)$, $\phi$ is an invariant state for $(\al_t)$.
Now our aim is to show that the reversible system
$(\cla,\al_t,\phi)$ satisfies (1.1) with $j_0$ as defined in
(2.4), for a suitable choice of $\IE_{0]}$. To that end, for any
element $\ul{x} \in \clm$, we verify by the relation
$<\ul{y},F_{t]}\ul{x}=<\ul{y},\ul{x}>$ for all $\ul{y} \in {\cal M}_{t]}$
that $$(F_{t]}\ul{x})_r=
\left \{ \begin{array}{lll} x_r,&\;\mbox{if}\;r < t;\\
\tau_{r_k-t}(...\tau_{r_{n-1}-r_{n-2}}(\tau_{r_n-r_{n-1}}(
x_{r_n})x_{r_{n-1}})...x_t),&\;\mbox{if}\;r=t\\
I,&\;\mbox{if}\;r > t \end{array} \right. $$
where $r_1 \le ..\le r_k \le t \le .. \le r_n$ is the support of
$\ul{x}$. We also claim that
\be
F_{s]}j^f_t(x)F_{s]}=j^f_s(\tau_{t-s}(x))\;\; \forall s \le t.
\ee
For that purpose we choose any two elements $\ul{y},\ul{y'} \in
\lambda({\cal M}_{s]})$ and check the following steps with the aid of
(2.2): $$ <\ul{y},F_{s]}j^f_t(x)F_{s]}\ul{y'}>=
<\ul{y},i_t(x)\ul{y'}>$$
$$=<\ul{y},i_s(\tau_{t-s}(x))\ul{y'})>.$$ Since $\lambda
(M_{s]})$ spans $\clh_{s]}$ it complete the proof of our claim.

\vsp
We also verify that $<z,V^*_+j^f_t(x)V_+y>_{\phi_0}=
\phi_0(z^*\tau_t(x)y)$, hence
\be
V^*_+j^f_t(x)V_+=\tau_t(x),\;\forall t \ge 0.
\ee

\vsp
We summarize this construction in the following theorem.

\vsp
\NI {\bf THEOREM 2.1:} There exists a Hilbert
space $\clh$ and a group of unitary operators $(S_t)$ with an
invariant vector $\Omega \in \clh$ so that
$$P^0_t=V^*_+S_tV_+,\;t \ge 0$$ and a triplet
$(\cla,\al_t,\phi)$ acting on $\clh$ so that the diagram (1.1)
commutes with the injective $*$ homomorphism $j^f_0$ as described
in (2.3) and the completely positive map $\IE_0(X)=V^*_+XV_+$.

\vsp
\newsection{ Asymptotic behavior of the stationary weak Markov 
process and the shift :}

\vsp
In this section we investigate how various properties (
ergodicity, weak mixing, strong mixing, etc ) of the system
$(\cla_0,\tau_t,\phi_0)$ is canonically related to that of the
minimal Markov shift $(\clh,S_t,F_{t]})$. To that end we first
introduce the following definition.

\bigskip
An element $y \in \cla_0$ is said to be {\bf invariant } for
$(\tau_t)$ if $\tau_t(y)=y$ for all $t \ge 0$. Thus any scaler
multiple of the identity is an invariant element. We say
$(\tau_t)$ is {\bf irreducible} if $ \tau_t(p) =p,\;t \ge 0 $
for a projection $p \in \cla_0$ implies that $p=0\;$ or $I$.

\smallskip
For each fixed $t \ge 0$, following Evans [Ev], we define
-conjugate linear maps $D_t: \cla_0 \times \cla_0 \rightarrow
\cla_0$ by  $D_t(y,y')=\tau_t(y^*y')
-\tau_t(y^*)\tau_t(y')$. Complete positiveness ( in fact
2-positive is enough ) of the map $\tau_t$ and $\tau_t(I)=I$
ensures that
\be
\tau_t(y^*)\tau_t(y) \le \tau_t(y^*y)\;\; \forall y \in \cla_0.
\ee
Thus (3.1) guarantees that $D_t$ is a non-negative -conjugate linear
form and a simple consequence of Cauchy-Schwartz inequality
says that for all $y' \in \cla_0,\;D_t(y,y')=0$ whenever
$D_t(y,y)=0.$ Now we conclude that $\tau_t(y^*y)=y^*y$ and $
\tau_t(y)=y$ if and only if $\tau_t(y'y)= \tau_t(y')y$ for all
$y' \in \cla_0$. The last statement in particular implies that
\be
\cln = \{ y : \tau_t(y)=y,\;\tau_t(y^*y) = y^*y,\;\tau_t(yy^*)=yy^*,\;t \ge
0 \}
\ee
is a $*$-subalgebra and for any projection $p \in \cln$, $\tau_t(px)=
p\tau_t(x)$. An element $p \in \cln$ is said to be irreducible if
there is no projection $q \in \cln$ such that $0 < q < p$. The 
following proposition is a simplification of Evan's [Ev] original work.

\vsp
\NI {\bf PROPOSITION \ 3.1} [Ev]: $(\tau_t)$ is irreducible if
and only if $\cln = \IC$.

\NI {\bf PROOF} : We show the non-trivial part of the
proposition. Let $y \in \cln$. Without loss of generality we
assume that $y^*=y$. From the relation
$\tau_t(y'y)=\tau_t(y')y\;\forall y' \in \cla_0$, we first note
that $\tau_t(y^n)=y^n$ for all $n \ge 1$ (by induction). Since
$\tau_t$ is a contraction on $\cla_0$ we get
$\tau_t(\psi(y))=\psi(y)$ for all bounded continuous real valued
functions on $\IR$. For a bounded Borel measurable function
$\psi$ we choose two family $\psi_n,\psi'_n $ of bounded
continuous functions so that $\psi_n,\psi'_n \raro \psi$
pointwise and $\psi_n \le \psi \le \psi'_n$. By positiveness of
$\tau_t$ we have $\psi_n(y) \le \tau_t(\psi(y)) \le \psi'_n(y)$
for all $n \ge 1$. Taking limit $n \raro \infty$ we conclude
that $\tau_t(\psi(y))=\psi(y)$ for all Borel measurable
functions $\psi$ on $\IR$. Since all invariant projections are
either $0$ or $1$, we conclude that the spectral family of $y$
are trivial. Hence $y$ is a constant multiple of the identity.
\qed

\bigskip
$\tau_t)$ is said to be {\it normal } if for each $t \ge 0$ the map $y
\rightarrow \tau_t(y)$ is {\it normal }, i.e. for any increasing net
$y_{\alpha},\;$ $\tau_t(lub\; y_{\alpha})\;= lub
\;\tau_t(y_{\alpha})$, where $lub$ denotes the least upper bound.
In such a case it is simple to check that ${\cal N}$ is $\sigma$-strong
closed and thus ${\cal N}$ is a von-Neumann algebra. A normal Markov
$(\tau_t,\;t \ge 0)$ semigroup on $\cla_0$ is said to be {\it weak$^{*}$ }
continuous if for each fixed $y \in
\cla_0$ the map $t \rightarrow \tau_t(y)$ is continuous with
respect to the $\sigma$-weak topology. In such a case there
exists a unique contractive semigroup $(\sigma_t)$ on the Banach
space of equivalence class of the trace class operators such that
$tr(\sigma_t(\rho)x)=tr(\rho\tau_t(x)) \forall t \ge 0,\;x \in \cla_0
,\; \rho \in (\cla_0)_*$. Now onwards we always assume $(\tau_t)$ is
weak$^*$ continuous.  At this point we note that unlike strong continuity 
on a Banach space, weak$^*$ continuity need not imply that the map $(t,y)
\rightarrow \tau_t(y)$ is jointly continuous, however the map
is sequentially jointly continuous i.e. $t_n \rightarrow t$ and 
$y_n \rightarrow y$ in the weak$^*$ topology then $\tau_{t_n}(y_n) 
\rightarrow \tau_t(y)$ in the weak$^*$ topology [AcM1], which serves 
our purpose for this exposition.

\vsp
A normal state $\phi_0$ is said to be {\it invariant} for $(\tau_t)$ if 
$\sigma_t(\phi_0)=\phi_0$ for all $t \ge 0$. It is well known that any 
Markov semigroup on a finite dimensional ${\cal A}_0$ admits an invariant
normal state. However for an infinite dimensional algebra ${\cal A}_0$,
a dynamical system may not admit an invariant 
normal state. Thus it remains an interesting open problem how to determine 
whether a given dynamical system admits a normal invariant state. In 
a series of papers, Fagnola and Rebolledo [FR1,FR2] addressed this problem 
when ${\cal A}_0={\cal B}({\cal H}_0)$ and found a sufficient condition 
which guarantees existence of an invariant normal state. In the following 
we also propose a simple criteria for existence of an invariant normal state 
which seems to be another sufficient condition for existence of an invariant 
normal state.

\vsp
\NI {\bf PROPOSITION 3.2 :} Let for a $\lambda > 0$, the resolvent 
$(R_{\lambda})(\rho)= \int^{\infty}_0e^{-\lambda t}\sigma_t(\rho)dt$ 
be a compact operator on the Banach space ${{\cal A}_0}_{*}$. Then 
$(\tau_t)$ admits a normal invariant state.
 
\vsp
\NI {\bf PROOF:} We fix any normal state $\phi$ on ${\cal A}_0$. Note
that the family $ \rho(t) = {1 \over t}\int^t_0\sigma_s(\rho)ds: t \ge 0$
is uniformly bounded. Thus by compactness of the resolvent we infer that
for any sequence $t_n \raro \infty$, there exists a
subsequence $t_{n_k}$ so that $ R_{\lambda}(\rho(t_{n_k})) $
converges in the Banach space norm topology. Now we use
the fact that $R_{\lambda}$
commutes with $(\sigma_t)$ to conclude that the limiting state is
an invariant state for $(\sigma_t)$. \qed

\smallskip
If $(\tau_t)$ admits an invariant faithful state, it follows
from (3.1) that $x^*x$ is an invariant element if $x$ is so. Thus
in such a case $\cln = \{ x \in \cla_0,\;
\tau_t(x)=x,\;t \ge 0 \}$ and there exists a 
norm one projection $E$ on $\cln$ so that weak$^*$ limit 
$_{\lambda  \raro 0}\; \lambda \int^\infty_0e^{-\lambda t}
\tau_t(x)dt=E(x)\; \forall \; x \in \cla_0$.  For more details we 
refer to Frigerio [Fr1]. In a recent paper Fagnola and Rebolledo [FR3],
investigated when an invariant normal state is faithful. In the following
we review their work and aim to prove an ergodic theorem for normal
invariant state. 

\vsp
Following [FR3] we now say a positive $x \in {\cal A}_0$ is  
{\it sub-harmonic} for $(\tau_t)$ if $\tau_t(x) \ge x \; \forall t 
\ge 0$. In such a case $\tau_t(x)$ is an increasing positive operator with
$\tau_t(x) \le ||x||1$, thus the strong limit $limit_{t \uparrow \infty}
\tau_t(x)$ exists and the limit is an invariant element for $(\tau_t)$.  
In the following we list few crucial property of sub-harmonic projection.

\vsp
\NI {\bf PROPOSITION 3.3: } Let $p$ be a sub-harmonic projection 
for $(\tau_t)$. Then the following hold:

\NI (a) for all $t \ge 0$, $p\tau_t(p) = \tau_t(p)p=p$.

\NI (b) $\tau_t(x(1-p))p=0$ for all $x \in \cla_0, t \ge 0$. 

\vsp
\NI {\bf PROOF: } For a quick verification for (a) we note 
that $p\tau_t(p)p \ge p$ and also $p(1-\tau_t(p))p \ge 0$. Thus we 
have $p(1-\tau_t(p))p= p$. Since $1-\tau_t(p) \ge 0$ we have $(1-\tau_t(p))p
=0$. For (b) we consider the non-negative conjugate bilinear form 
$\psi(x^*_1\tau_t(x^*_2y_2)y_1)$ for a 
positive normal state and use once more Cauchy-Schwartz in-equality to 
conclude that $\psi(x^*_1\tau_t((x^*_2(1-p))p)=0$ for any $x_1,x_2 \in \cla_0$
and $t \ge 0$. \qed 

\vsp
For a projection $p$, $\cla^p_0=p\cla_0p$ is a von-Neumann acting on the Hilbert 
subspace $p\clh_0$. Thus for a sub-harmonic projection $p$ we verify by 
Proposition 3.3 that $(\tau^p_t)$ defined by $\tau^p_t(x)= p\tau_t(x)p,\; 
x \in \cla^p_0$ is a Markov semigroup. Let the strong limit 
$\tau_t(p) \uparrow y$ as $t \uparrow \infty$. By Proposition 3.3. 
(a) we have $py=yp=p, p \le y \le 1$ and $\tau_t(y)=y \forall t \ge 0$. 
Thus $p\tau_t(1-y^2)p=p\tau_t(p(1-y^2)p)p=0$.
So we also have $p\tau_t(y^2)=\tau_t(y^2)p=p$ for all $t \ge 0$.
Since $\tau_t(p) \le \tau_t(y^2) \le \tau_t(y)=y$, the strong limit of 
$\tau_t(y^2)$ as $t \raro \infty$ is also $y$. In case $y^2$ 
is also an invariant element for $(\tau_t)$, we have $y^2=y$. In general 
$y^2$ need not be an invariant element even for an irreducible classical 
Markov semigroup $(\tau_t)$. In general $y^2$
need not be an invariant element even for an irreducible $(\tau_t)$.
We give a simple counter example in classical Markov chain in the
following.
Consider three state discreet time Markov chain where
two of it's states are absorbing and third state is a transient one
with equal transition probability $ {1 \over 2}$ to those two absorbing
state. The chain is irreducible in
the sense of [Ev]. Indicator function of an absorbing state is
a sub-harmonic function for which $y=(1,0,{1 \over 2})$ or 
$y=(0,1,{1 \over 2})$ depending on which indicator function we 
have taken as $p$. 
 
\vsp
\NI {\bf PROPOSITION 3.4: } Let $p$ be a sub-normal projection and 
$y=\mbox{s.lim}\tau_t(p)$. Then for any $z \in \clb(\clh_0)$ the following 
statements are equivalent:

\NI (a) $yz=0$

\NI (b) $\tau_t(p)z=0$ for all $t \ge 0$.
                               
\vsp
\NI {\bf PROOF:} That (b) implies (a) is 
obvious. For the converse, note that $z^*\tau_t(p)z \le zyz=0$
by (a), hence (c) follows. \qed 

\vsp
In case $(\tau_t)$ is the semigroup associated with a quantum mechanical 
Fokker-Planck equation (see the last section), we will explore 
this explicit criteria further. In the following we will 
investigate its implication. In case $y=1$ by Cauchy-Schwartz inequality 
$|\psi(\tau_t((1-p)x)|^2 \le \psi(\tau_t(1-p) \tau_t(1-p))\psi(x^*x)$ for 
a normal state $\psi$, we conclude that $\tau_t((1-p)x) \raro 0$ in the 
weak$^*$ topology as $t \raro \infty$ for all $x \in {\cal A}_0$.  

\vsp
We recall an interesting result from Fagnola-Rebolledo [FR3] in the 
following proposition.

\vsp
\NI {\bf PROPOSITION 3.5 :}[FR3] Let $\phi_0$ be an invariant normal state
on ${\cal A}_0$. Let $p$ be the support of $\phi_0$. Then $p$ is 
sub-harmonic. 

\NI {\bf PROOF :} Since $\phi_0(p(1-\tau_t(p))p)=1-1=0$ and $p$ is the 
minimal projection we have $p(1-\tau_t(p))p=0$. Since $1-\tau_t(p) \ge 0$,
we conclude that $(1-\tau_t(p))p=0$. Hence $\tau_t(p)=p + 
p^{\perp}\tau_t(p)p^{\perp} \ge p$. \qed

%%%%%%%%%%%%%%%%%%%%%%%%%%%%%%%%%%%%%%%%%%%%%%%%%%%%%%%%%%%%%%%%%%%%%%%%%%%%%
\vsp
The following result shows that faithfulness of the normal invariant 
state can be removed for an von-Neumann-Frigerio type of ergodic theorem. 

\vsp
\NI {\bf THEOREM 3.6: } Let $\phi_0$ be an invariant normal state
for $(\tau_t)$ which has support $p$ so that the strong limit 
$\uparrow \tau_t(p) = 1$ as $t \uparrow \infty$. Then the following
statements are equivalent:

\NI (a) $\{ pxp : p\tau_t(pxp)p=pxp \}= \{zp:\;z \in \IC \}$

\NI (b) for all $x \in \cla_0$, $\lambda \int e^{- \lambda t}p\tau_t(pxp)p 
\raro \phi_0(x)p $ in the weak$^*$ topology as $\lambda \raro 0$. 

\NI (c) for all $x \in \cla_0$, $ \lambda \int e^{-\lambda t}\tau_t(x)dt 
\raro \phi_0(x)1$ in the weak$^*$ topology as $\lambda \raro 0$. 

\vsp
\NI {\bf PROOF:} Since $\phi_0$ restricted to $\cla^p_0$ is faithful, 
equivalence of (a) and (b) follows by Theorem 2.1 in [Fr1]. That (c) 
implies (b) is trivial. We are left to show that (c) implies (b). 
Since $\tau_t((1-p)x) \raro 0$ as $t \raro \infty$, we 
need to verify (c) for elements in $\cla^p_0$ only. To that end 
first note that $\tau_{t+s}(pxp)
= \tau_t(p\tau_s(pxp)p) + \tau_t(p\tau_s(pxp)p^{\perp})$$ 
$$ + \tau_t(p^{\perp} \tau_s(pxp) p) + 
\tau_t(p^{\perp} \tau_s(pxp)p^{\perp})$
and limsup$_{\lambda \raro 0}
|\psi(\lambda \int e^{-\lambda s}\tau_{t+s}(x)ds|$ is independent of $t$ that
we choose. On the other hand limit$_{t \raro \infty}$
limsup$_{\lambda \raro 0} |\psi (\tau_t(z \lambda R_{\lambda} (pxp)p^{\perp})|
\le limit_{t \raro \infty} ||x||\;||z|| \psi(\tau_t(p^{\perp})$
is zero for any $z \in \cla_0$. Hence (c) follows once we use 
(b) for $pxp$ with $\phi_0(x)=0$. The general 
result follows once we verify $\lambda \int e^{-\lambda t}\tau_t(p)dt \raro 
1$ as $\lambda \raro 0$ by our hypothesis $\tau_t(p) \raro 1$. \qed       

\vsp
One more important point we note that for a sub-normal projection $p$ for $(\tau_t)$,
if the reduced dynamical system admits a normal invariant state $\phi_0$ on $\cla^p_0$, then
we can extend (need note be unique) the state to be an invariant normal state for 
the entire dynamics by $\phi_0(x)= \phi_0(pxp)$. However in case $\mbox{s.limit}_{t 
\raro \infty}\tau_t(p)=1$ and the reduced dynamics $(\tau_t^p)$ admits a 
faithful normal ergodic state then the extension to $\cla$ is unique. The 
conditions $\tau_t(p) \uparrow 1$ is also necessary for ergodicity for the
entire system.     

\vsp
Now we fix a normal Markov semigroup $(\tau_t)$ on $\cla_0$ which admits 
a normal invariant state and consider the Markov shift 
$(S_t)$ constructed on the minimal Hilbert space $\clh$ in Section 2. 
$(S_t)$ is strongly continuous once $(\tau_t)$ is continuous in the weak$^*$ 
topology. Converse is also true provided $\phi_0$ is faithful. For details 
we refer to [AcM2]. 

\vsp
\NI {\bf PROPOSITION  3.7 :} [AcM2] Let $(\tau_t)$ be weak$^*$ continuous 
with a normal invariant state $\phi_0$. Then the following statements are 
equivalent:

\smallskip
\noindent (a) limit$_{\lambda \raro 0}\int e^{-\lambda t}\phi_0(y\tau_t(x))dt
= \phi_0(x)1$ for all $x$ and $y \in \cla_0$;

\smallskip
\noindent (b) $(S_t)$ is ergodic, i.e. $\{f:\;S_tf=f\;\forall t
\in \IR \} = \IC \Omega.$

\bigskip
$\underline{ PROOF: }$ We refer once more to [AcM2] for a proof.

\vsp
We recall few more results from [AcM2] in the following proposition:

\vsp
\NI {\bf PROPOSITION 3.8:} Let $(\tau_t)$ be $\sigma-$ weakly 
continuous dynamical semigroup with a normal invariant state $\phi_0$ 
and $(\clh,S_t)$ is the minimal Markov shift.

\NI (i) The following statements are equivalent:

\NI (a) For all $h_1,h_2 \in \clh,\;$ limit$_{T \raro \infty }{1
\over T}\int^T_0|<h_1,S_th_2>-<h_1,\Omega><\Omega,h_2>|dt\;=0$;

\NI (b) The spectrum of $(S_t)$ in the orthocomplement of
$\IC\Omega$ is continuous.

\NI (c) For all $x,y \in \cla_0,\;$ limit$_{T \raro \infty }{1
\over T}\int^T_0|\phi_0(x\tau_t(y))-\phi_0(x)\phi_0(y)|dt\;=0$;

\NI (ii) The following statements are equivalent:

\NI (d) For all $h_1,h_2 \in \clh,\;$ limit$_{t \raro \infty
}<h_1,S_th_2>=<h_1,\Omega><\Omega,h_2>$;

\NI (e) For all $x,y \in \cla_0,\;$ limit$_{t \raro \infty
}\phi_0(x\tau_t(y))=\phi_0(x)\phi_0(y)$. 

\NI (f) For any $x \in \cla_0$, weak$^*$ limit of $j_t(x) \raro \phi_0(x)$
as $t \raro \infty$.

\NI {\bf PROOF: } For the proof once more we refer to [AcM2]. 

\vsp
We say $(\cla_0,\tau_t,\phi_0)$ is {\bf weak mixing } if (a)
holds and {\bf strong mixing } if (d) holds. It is obvious that
weak mixing implies ergodicity and strong mixing implies weak
mixing. A simple consequence of the spectral theorem and
Riemann-Lebesgue lemma implies strong mixing whenever the
spectrum of $(S_t)$ in the orthocomplement of $\IC\Omega$ is
absolutely continuous. In general, it is rather hard to find a
useful criteria for absolute continuity of the spectrum in the
orthocomplement of $\IC \Omega$. On the other hand, it is
still not clear even in the classical case whether, this is also
necessary [Pa]. At this point we also note that if $\phi_0$ is also
faithful, strong mixing guarantees that weak$^*$ limit of $\tau_t(x) \raro
\phi_0(x)$ as $t \raro \infty$ for all $x \in \cla_0$. We postpone this
issue now.  

\vsp
Since $F_{t]}=j_t(I)$, $I-F_{t]} \in \cla$ and
$(I-F_{t]})\Omega=0$, and thus $\Omega$ is not a separating
vector for $\cla$ even if $\phi_0$ is so. So the support of the
state $\phi$ on $\cla$ is a proper projection $P_{\phi} \in
\cla$ defined by $$P_{\phi}=[\cla'\Omega]$$ where $\cla'$ is the
commutant of $\cla$. Since $(\al_t)$ preserves $\cla$, we check
also that $(\al_t)$ preserves $\cla'$ and thus $P_{\phi}$ is an
invariant element for $\al_t$. Since $F_{t]}\Omega=\Omega$ and
$F_{t]} \in \cla$, we check also that
$P_{\phi}F_{t]}X'\Omega=F_{t]}P_{\phi}X'\Omega=X'\Omega$. In
other-words $P_{\phi}F_{t]}=F_{t]}P_{\phi}=P_{\phi},\;t \in \IR.$
Motivated by the well known notion, {\it \bf Kolmogorov shift }, in ergodic
theory, we introduce the following notion.

\vsp
We say the minimal forward weak Markov process 
$(\clh,j_t,F_{t]},S_t,\Omega)$ associated with
$(\cla_0,\tau_t,\phi_0)$ is having {\bf Kolmogorov's property} on 
$\clh$ if $\cap_{t \in \IR}
F_{t]}=\IC\Omega.$ It is obvious that Kolmogorov's property implies that
$P_{\phi}=|\Omega><\Omega|$. In such a case $\cla=\clb(\clh)$. The following
important proposition gives a criteria for Kolmogorov's property.

\vsp
\NI {\bf THEOREM 3.9:} $F_{t]} \rightarrow |\Omega><\Omega|$ as $ 
t \raro -\infty$ if and only if
$$\mbox{lim}_{t \raro \infty } \phi_0(\tau_t(x)\tau_t(y)) = \phi_0(x)\phi_0(y)
\;\forall x,y \in \cla_0.$$ In such a case the following hold:

\NI (a) $j_t(x) \raro \phi_0(x)|\Omega><\Omega| \;\forall x \in \cla_0$ 
in the weak$^*$ topology as $t \raro -\infty$.  

\NI (b) $\alpha_t(X) \raro \phi(X)$ as $t \raro \infty$ in the weak$^*$ 
topology as $t \infty -\infty$ for all $X \in \bigcup_{s \in \IR} 
{\cal A}_{s]}$.  

\vsp
\NI {\bf PROOF:} We need to show that $\mbox{lim}_{t \raro -\infty }
F_{t]}=|\Omega><\Omega|$ if and only if $\mbox{lim}_{t \raro \infty }
\phi_0(\tau_t(x)\tau_t(y)) = \phi_0(x)\phi_0(y)\;\forall x,y \in \cla_0.$
Since the family $F_{t]}$ is uniformly norm bounded, $\mbox{lim}_{
t \raro -\infty }
F_{t]}=|\Omega><\Omega| $ if and only if
$$\mbox{lim}_{t \raro -\infty } <\ul{x},F_{t]}\ul{y}>=<\ul{x},\Omega><\Omega,
\ul{y}>.$$
The result follows once we note that for any fix $\ul{x},\ul{y} \in \clh$
if $t \le r_1,r'_1$, where
$r_1,r'_1$ are the lowest support of $\ul{x}$ and $\ul{y}$ respectively,
$$<\ul{x},F_{t]}\ul{y}> = <F_{t]}\ul{x},F_{t]}\ul{y}>$$
$$=
\phi_0[ (\tau_{r_1-t}(...\tau_{r_{n-1}-r_{n-2}}(\tau_
{r_n-r_{n-1}}(x_{r_n})x_{r_{n-1}})...x_{r_1}) )^*
\;$$
$$\tau_{r'_1-t}(...\tau_{r'_{m-1}-r'_{m-2}}(\tau_
{r'_m-r'_{m-1}}(y_{r'_m})y_{r'_{m-1}})...y_{r'_1})]$$
For (a) we also note that $<\ul{x},j_t(z)\ul{y}>$
$$=\phi_0[ (\tau_{r_1-t}(...\tau_{r_{n-1}-r_{n-2}}(\tau_
{r_n-r_{n-1}}(x_{r_n})x_{r_{n-1}})...x_{r_1}) )^*
\;$$
$$z \tau_{r'_1-t}(...\tau_{r'_{m-1}-r'_{m-2}}(\tau_
{r'_m-r'_{m-1}}(y_{r'_m})y_{r'_{m-1}})...y_{r'_1})]$$ 
and for any $x,y,z \in \cla_0$ 
$\phi_0(\tau_t(x)z\tau_t(y)) \raro \phi_0(x)\phi_0(z)\phi_0(y)$ as $ t \raro
\infty$. For (b) we claim that
$\bigcap_{s \in \IR}{\cal A}_{s]}$ is von-Neumann algebra generated by
$|\Omega><\Omega|$. Our claim follows since for any such $X$ we have $
F_{t]}XF_{t]}=X$ for all $ t \in \IR$. Thus by taking limit $t \raro 
-\infty$ we have $X= \phi(X)|\Omega><\Omega|$. 
We recall that for any $s,t \in \IR$ $\alpha_{t}({\cal A}_{s]})=
{\cal A}_{s+t]}$. Hence once we fix any $X \in {\cal A}_{s}$ by 
weak$^*$ compactness of the unit ball in ${\cal A}$ we conclude 
that the limit points as $t$ diverges to $\infty$ is equal to 
$\phi(X)|\Omega><\Omega$. Since the limit point is uniquely determined,
the result follows. This complete the proof \qed

\vsp
One interesting feature appears in Theorem 3.8 that the two point
correlation $\phi(\alpha_t(X)Y) \raro \phi(X)\phi(Y)$ as long as
$Y$ is an element in one of the local von-Neumann algebras. This
asymptotic abelianess holds good to the $C^*$ algebra completion of
the $*$ algebra $\bigcup_{t \in \IR} {\cal A}_{t]}$. Since 
${\cal A}={\cal B}({\cal H})$ and spectrum of $H$ contains $\IR$
we conclude that the $C^*$ algebra is strictly contained in ${\cal A}$
and asymptotic abelianess do not hold for ${\cal A}$.

%%%%%%%%%%%%%%%%%%%%%%%%%%%%%%%%%%%%%%%%%%%%%%%%%%%%%%%%%%%%%%%%%%%%%%%%%%%%%%%%\vsp
By polarization identity, we check that $(\tau_t)$ is a K-shift if
and only if lim$_{t \raro \infty}||P^0_tx||=\phi_0(\tau_t(x^*)\tau_t
(x)) \raro 0 $ for all $x \in \cla_0$ such that $\phi_0(x)=0$.
Before we start investigating this criteria further we note once 
more by Cauchy-Schwartz in-equality that $(\tau_t)$ is strong mixing and 
$\{x \in \cla_0:\tau_t(x^*)\tau_t(x)=\tau_t(x^*x), \forall t \ge 0
\}$ is trivial i.e. $\{ \lambda 1,\;\lambda \in \IC \}$ whenever 
$(\clh,F_{t]},S_t)$ is a K-shift.

\vsp
The following result shows why we need infinite dimensional Hilbert
space in order to construct a strong mixing dynamical system which 
is not a K-shift. 

\vsp
\NI {\bf COROLLARY 3.10:} Let the resolvent $R_{\lambda}=\int^{\infty}_0
e^{-\lambda t} P_t dt, \lambda > 0,$ be a compact operator for some 
$\lambda > 0$. Then strong mixing Markov shift is also a K-shift.

\vsp
Now inspired by Frigerio's seminal paper [Fr1], we revisit his work and 
find a sufficient condition for strong mixing. We set 
\be
\clf=\{x:\tau_t(x^*)\tau_t(x)=\tau_t(x^*x);
\;\tau_t(x)\tau_t(x^*)=\tau_t(xx^*),\;\forall t \ge 0 \}
\ee
We claim that 
$\clf$ is a von-Neumann sub-algebra. To that end first we note by $2-$ positive $((\tau_t(x_i^*x_j))) \ge (( \tau_t(x_i^*)\tau_t(x_j) ))$ where $x_i:i=1,2$
are any elements in $\cla$. Thus for any $t \ge 0$, by choosing $x_1=x$ and $x_2=y$ we conclude  
that 
\be
\tau_t(x^*y)=\tau_t(x^*)\tau_t(y)\;\;  \mbox{whenever}\;\; \tau_t(x^*x)=\tau_t(x^*)\tau_t(x) 
\ee
Now it is a routine work to check that $\clf$ is a linear space and a $*$-algebra. That it is a von-Neumann algebra follows by the normality of $(\tau_t)$ 
and (3.4).

\vsp
\NI {\bf PROPOSITION 3.11: } [Fr1] Let $\phi_0$ be a faithful normal 
invariant state for $(\tau_t)$.  
If $\cln=\clf$ then weak$^*- \mbox{limit}
_{t \raro \infty } \tau_t(x)=E(x)\;\; \forall x \in \cla$, where $E$ is
the unique norm one projection on $\cln$. 

\vsp
\NI {\bf PROOF :} For any $x,y \in \cla$, $\mbox{limit}_{t \raro \infty}
\phi_0(D_t(x,y))=\mbox{limit}_{t \raro \infty}<x\omega,(I-P^*_tP_t)y\omega>
=
<x\omega,Qy\omega> $, where $Q=s-\mbox{lim}_{t \raro \infty}I- P_t^*P_t$. 
Thus $\phi_0(D_t(\tau_s(x),\tau_s(y)=\phi_0(D_{s}(x,y))-\phi_0(D_{s+t}(x,y))
\raro 0$ as $s \raro \infty$. By Cauchy-Schwartz
inequality we conclude that $|\phi_0(D_t(\tau_s(x),y))|^2 \le \phi_0(D_t(\tau_s(x),\tau_s(x))) \phi_0(D_t(y,y)) \raro 0$ as $s \raro
\infty$. Thus for any weak$^*$ limit point $x_{\infty}$ as $s \raro \infty$ of the
norm bounded net $\{\tau_s(x)\}_s$, we have $D_t(x_{\infty},x_{\infty})=0$. 
Since $x_{\infty}^*$ is also a limit point of the norm bounded net $\{ 
\tau_s(x^*)\}_s$, we conclude that $D_t(x^*_{\infty},x^*_{\infty})=0$ for
all $t \ge 0$. Thus $x_{\infty} \in \clf$. Since $E$ is a norm one
projection on $\cln$, 
we have $\tau_s(x)=E(x)+(I-E)(\tau_s(x)) \forall s \ge 0.$ Thus any limiting
point $x_{\infty}$ satisfies  
$x_{\infty}=E(x)+(I-E)(x_{\infty})$. If $\cln=\clf$, $(I-E)(x_{\infty})=0$, 
so $x_{\infty}=E(x)$, which is uniquely determined. Since this holds for any weak$^*$ limit point, 
the result follows by weak$^*$ compactness of the unit ball of $\cla$. 

\vsp
The following theorem suggest that we can have steady state which need not 
be faithful. 

\vsp
\NI {\bf THEOREM 3.12:} Let $\phi_0$ be a normal invariant state for
$(\tau_t)$ and $p$ be the support projection for $\phi_0$ so that 
the strong limit of $\tau_t(p) \uparrow 1$ as $t \uparrow \infty$. 
Then the following are equivalent:

\NI (a) $\{ pxp: p\tau_t(px^*p)p\tau_t(pxp)p= p\tau_t(px^*pxp)p,\;t 
\ge 0 \} = \{ \lambda p:\;\lambda \in \IC \}$ 
 
\NI (b) for all $x \in \cla_0,$ $\tau_t(x) \raro \phi_0(x)1$ in the 
weak$^*$ topology.

\vsp
\NI {\bf PROOF: } Since $\tau_t((1-p)x) \raro 0$ 
in the weak$^*$ topology, it is good enough if we verify that (a) is 
equivalent to $\tau_t(pxp) \raro \phi_0(x)1$ in the weak$^*$ topology 
as $t \raro \infty$. To that end we first note that limsup$_{t \raro 
\infty} \psi(\tau_{s+t}(x))$ is independent of $s \ge 0$ we choose. On the 
other hand we write $\tau_{s+t}(pxp) = \tau_s(p\tau_t(pxp)p) + 
\tau_s(p\tau_t(pxp)p^{\perp})+ \tau_t(p^{\perp}\tau_s(pxp)p) +
\tau_s(p^{\perp}\tau_t(pxp)p^{\perp})$ and use the fact for any 
normal state $\psi$ we have 
$\mbox{limsup}_{t \raro \infty}|\psi(\tau_s(z\tau_t(pxp)p)| 
\le ||x||\;||z||\;|\psi(\tau_s(p))| $ for all $z,x \in \cla_0$. 
Thus by our hypothesis on the support and Proposition 3.11 we conclude 
that $\mbox{limsup}_{t \raro \infty}|\psi(\tau_t(pxp))|= 0$ for 
all $x$ for which $\phi_0(x)=0$. For the general case, we use 
the identity $\psi(\tau_t(pxp))=\psi(\tau_t(p(x-\phi_0(x))p) + 
\phi_0(x)\psi(\tau_t(p))$ and our hypothesis $\tau_t(p) \uparrow 1$ as 
$t \raro \infty$.  \qed

\newsection{ Time reverse weak Markov process and 
Quantum detailed balance :}

Following [AcM], we will consider the time reverse process associated 
with the KMS-adjoint ( Or Petz adjoint ) quantum dynamical semigroup 
$(\cla,\tilde{\tau}_t,\phi_0)$. We aim to investigate how far 
various properties of the dynamical semigroup are time reversible. First 
we recall 
from [AcM] time reverse process associated with the KMS-adjoint 
(Petz-adjoint ) semigroup in the following paragraph.

\vsp
Let $\phi_0$ be a faithful state and 
without loss
of generality let also $(\cla_0,\phi_0)$ be in the standard form
$(\cla_0,J,{\cal P},\omega_0)$
[BrR] where $\omega_0
\in \clh_0$, a cyclic and separating vector for $\cla_0$, so
that $\phi_0(x)= <\omega_0,x\omega_0>$ and the closer of the
close-able operator $S_0:x\omega_0
\raro x^*\omega_0, S$ possesses a polar decomposition
$S=J\Delta^{1/2}$ with the self-dual positive
cone $\clp$ as the closure of $\{ JxJx\omega_0:x \in \cla_0 \}$ in
$\clh_0$. Tomita's [BrR] theorem says that
$\Delta^{it}\cla_0\Delta^{-it}=\cla_0,\;t
\in \IR$ and $J\cla_0J=\cla'_0$, where $\cla'_0$ is the
commutant of $\cla_0$. We define the modular automorphism group
$\sigma=(\sigma_t,\;t \in \IR )$ on $\cla_0$
by
$$\sigma_t(x)=\Delta^{it}x\Delta^{-it}.$$
Furthermore for any normal state $\psi$ on $\cla_0$
there exists a unique vector $\zeta \in {\cal P}$ so that $\psi(x)=
<\zeta,x\zeta>$.

We consider the unique Markov semigroup $(\tau'_t)$ on the commutant
$\cla'_0$ of $\cla_0$ so that $\phi(\tau_t(x)y)=\phi(x\tau'_t(y))$ for all
$x \in \cla_0$ and $y \in \cla'_0$. 
We define  
weak$^*$ continuous
Markov semigroup $(\tilde{\tau}_t)$ on $\cla_0$ by $\tilde{\tau}_t(x)=J\tau'_t(JxJ)J.$
Thus we have the following adjoint relation
\be
\phi_0(\sigma_{1/2}(x)\tau_t(y))=\phi_0(\tilde{\tau}_t(x)\sigma_{-1/2}(y))
\ee
for all $x,y \in \cla_0$, analytic elements for $(\sigma_t)$. One
can as well describe the adjoint semigroup as Hilbert space adjoint of
a one parameter contractive semigroup $(P_t)$ on a 
Hilbert space defined by $P_t:\Delta^{1/4}x\omega_0=\Delta^{1/4}\tau_t(x)\omega_0.$ For more details we refer to [Ci].   

\vsp
Once $\phi_0$ is also faithful, there exists also a unique 
backward weak
Markov 
process $(j^b_t)$ which generalizes Tomita's representation 
and a family of projections $F_{[t}:\; t \in \IR$ so that 
$$F_{[s}j^b_t(x)F_{[s}=j^b_s(\tilde{\tau}_{s-t}(x))$$ 
for $-\infty < t \le s < \infty $. For more details and
the following result we refer to [AcM].

\vsp
\NI {\bf THEOREM 4.1: }[AcM] We consider the weak Markov processes $(\cla,
\clh,F_{t]},F_{[t},S_t,j^f_t,\;j^b_t\;\;t \in \IR,\; \Omega)$ associated with $(\cla_0,\tau_t,
\;t \ge 0,\;\phi_0)$
and the weak Markov processes $(\tilde{\cla},\tilde{\clh},\tilde{F}_{t]},
\tilde{F}_{[t},\tilde{S}_t,\;\tilde{j}^f_t,\;\tilde{j}^b_t,\; t \in \IR,\; \tilde{\Omega})$ associated with $(\cla_0,\tilde{\tau}_t,\; t \ge 0,\;\phi_0)$. There exists an unique
anti-unitary operator
$U_0:\clh \raro \tilde{\clh}$ so that 

\NI (a) $U_0 \Omega = \tilde{\Omega}$;

\NI (b) $U_0 S_t U^*_0 = \tilde{S}_{-t}$ for all $t \in \IR$;

\NI (c) $U_0 j^f_t(x) U_0 = \tilde{j}^b_{-t}(x),\;U_0J^b_t(x)U_0=\tilde{j}^f_{-t}(x)$ for all $t \in \IR$;

\NI (d) $U_0F_{t]}U^*_0=\tilde{F}_{[-t},\;\;U_0F_{[t}U^*_0=\tilde{F}_{-t]}$ 
for all $t \in \IR$;

\vsp
A simple corollary of Theorem 4.1 is the following result.

\vsp
\NI {\bf COROLLARY 4.2:} $(\tilde{\clh},\tilde{S}_t,\tilde{F}_{t]})$ is a 
K-shift if and only if $\phi_0(\tilde{\tau}_t(x)\tilde{\tau}_t(y)) \raro \phi_0(x)\phi_0(y)$ as $t \raro \infty$.
In other words $\cap_{t \in \IR} F_{[t}=|\Omega><\Omega|$ if and only if
$\phi_0(\tilde{\tau}_t(x)\tilde{\tau}_t(y)) \raro \phi_0(x)\phi_0(y)$ as 
$t \raro \infty$. 

\vsp 
So it is now simple to verify directly by Theorem 4.1 that ergodicity, weak-mixing, strong mixing are
time reversible. It is not very transparent whether the same fact holds 
also for
K-shift property.
In the classical case [Pa] this property is well known to be 
equivalent to strictly
positive
 {\it dynamical entropy } $h(\theta,\zeta)$ of the shift $\theta$ 
for any non-trivial partition
$\zeta$ of the measure space. Since $h(\theta,\zeta)=h(\theta^{-1},\zeta)$
we conclude that 
K-shift property is also time reversible. However such a notion and result 
in the
general case is still missing [OhP]. We conjecture the following.  

\vsp
\NI {\bf CONJECTURE 4.3:} $\cap_{t \in \IR} F_{t]}=|\Omega><\Omega|$ if and 
only if $\cap_{t \in \IR} F_{[t} =|\Omega><\Omega|$. 

\vsp
We will verify this conjecture with an affirmative answer when $\cla_0=
\clb(\clh_0)$, algebra of all bounded operators, more general case will
include type-I von-Neumann algebra with center completely atomic. 

\vsp
Before
we proceed we find an alternative criteria for strong mixing in the 
following theorem. To that end we introduce $\clg=\{x \in \cla_0
:\tilde{\tau}_t\tau_t(x)=x, \; t \ge 0 \}.$

\vsp
\NI {\bf THEOREM 4.4:} Let $(\cla_0,\tau_t,\phi_0)$ is a quantum dynamical
system with $\phi_0$, a faithful normal invariant state for $(\tau_t)$. 
If $\cln = \clg$
then weak$^*$ 
limit$_{t \raro \infty }\tau_t(x)=E(x)$.

\vsp
\NI {\bf PROOF :} In spirit proof is similar to that of Theorem 3.8. 
We consider the bilinear form $d_t(x,y)=\phi_0(x^*JyJ)- 
\phi_0(\tau_t(x^*)J\tau_t(y)J)
\; t \ge 0$. That $d_t(x,x) \ge 0$ follows from the unital positive property of
$\tau_t$. 
Also note that                                                                  $d_t(\tau_s(x),\tau_s(y))=d_s(x,y)\;-\;d_{s+t}(x,y)$.
Thus $d_t(x,x)$ is monotonically increasing and bounded above by $\phi_0(Jx^*Jx)$.
So along the line of 
Proposition 3.8 we conclude that any weak$^*$ limit point of the net 
$\{ \tau_t(x) \}$ as $t \raro \infty$ will be an element say $x_{\infty}$ 
satisfying $d_t(x_{\infty}
,x_{\infty})=0$ for all 
$t \ge 0$. By Cauchy-Schwartz inequality $d_t(x,x)=0$ if and only if $
d_t(x,y)=0$ for all $y \in \cla$ i.e. $\phi_0(\tilde{\tau}_t\tau_t(x)JyJ)=\phi_0(xJyJ)$ for all $y \in \cla_0$. Thus $x_{\infty} \in \clg$  
which
is same as $\cln$ by our hypothesis. Since $\tau_t(x)=E(\tau_t(x))+(I-E)(\tau_t(x))$ and $E(\tau_t(x))=\tau_t(E(x))=E(x)$ we conclude that $x_{\infty}=E(x)$.
Thus the result follows from weak$^*$ compactness of the unit ball. \qed

\vsp
One natural question whether the sufficient condition in Theorem 4.4 is really 
different 
from Frigerio's criteria. By (3.1) we note that $||\tilde{\tau}_t\tau_t(x)\omega_0|| \le ||\tau_t(x)\omega_0||$ for all $x \in \cla_0$, thus $\clg \sbs \clf$.
Since $\cln \sbs \clg$ we conclude that Frigerio's condition $\clf=\cln$ also
guarantees that $\clg=\cln$. It is not clear whether the reverse inclusion
is true. However, in case modular automorphism commutes with the Markov 
semigroup
then [Fr2] $\phi_0(\tau_t(x)\tau_t(y))=\phi_0(x\tilde{\tau}_t\tau_t(y))$ for
any $x,y  \in \cla_0$, hence $\clf=\clg$. Since strong mixing property is
time reversible, we also get sufficient conditions $\tilde{\clf}=\cln$ or
$\tilde{\clg}=\cln$ ( note that $\tilde{\cln}=\cln$) associated with the 
adjoint Markov semigroup $(\tilde{\tau}_t)$ for strong mixing. Once more 
it is not resolved whether $\clf=\tilde{\clf}$ or $\clg=\tilde{\clg}$. However
the following proposition indicates that they are essentially same. To that
end we introduce $\clg_s=\{x:\;\tilde{\tau}_t(\tau_t(x))=x,\;s \ge t \ge 0 \}$
and $\clf_s=\{x:\;\;\tau_t(x^*)\tau_t(x)=\tau_t(x^*x),\; \tau_t(x)\tau_t(x^*)=
\tau_t(xx^*),0 \le t \le s \}$ for each $s > 0$.

\vsp
\NI {\bf PROPOSITION 4.5: } For each $0 < s,\; \clg_s=\cln$ if and only if
for each $0 < s,\; \tilde{\clg}_s=\cln$.  
Same hold for $\clf$. 

\vsp
\NI {\bf PROOF: } Since $\cln \sbs \clg_s$, we only need to show $\tilde{\clg}_s
\sbs \cln$ if $\clg_s = \cln$ for each $s > 0$. So we fix $s > 0$ and let 
$x \in \tilde{\clg}_s$. So $\tilde{\tau}_t\tau_t(x)=x,\;0 \le t \le s.$ 
Hence $\tau_t\tilde{\tau}_t(y)=y$ where $y=\tau_t(x)$. Since 
$\phi_0(\tilde{\tau}_t(z)J\tilde{\tau}_t(z)J)$ is a monotonically decreasing 
function for any $z \in \cla_0$ we conclude that $\phi_0(\tilde{\tau}_r(y)
J\tilde{\tau}_r(y))=\phi_0(yJyJ)$ for $0 \le r \le t$, thus we have $ 
\tau_t(x) \in \clg_t$. Thus we have $\tau_t(x) \in \cln$ for all $t > 0$. 
Now taking limit $t \raro 0$, we conclude the required result. We omit
the proof for $\clf$.  \qed

\vsp
\NI {\bf PROPOSITION 4.6: } Let $(\cla_0,\tau_t,\phi_0)$ be a quantum dynamical semigroup with 
a faithful normal state $\phi_0$. Then  
$\phi_0(\tau_t(x)\tau_t(y)) \raro \phi_0(x)\phi_0(y) \forall \; x,y \in \cla_0
$ as
$t \raro \infty$ if and only if $\phi_0(J\tau_t(x)J\tau_t(y)) \raro \phi_0(x)\phi_0(y) \; \forall x,y \in \cla_0$ as
$t \raro \infty$. 

\vsp
\NI {\bf PROOF:} Since $J$ is an anti-unitary operator, in particular contraction, thus `if part' is obvious. For the converse statement, 
first note that $\Delta^{1/4}|\tau_t(x)\omega> \raro 
\phi_0(x) |\omega>$ strongly as $t \raro \infty$. Now we use the fact that
$\Delta$ is a closed operator to
conclude that $J\Delta^{1/2}|\tau_t(x)\omega> \raro \overline{\phi_0(x)}
|\omega>|$ strongly as $t \raro \infty$. But $J\Delta^{1/2} |x\omega=
|x^*\omega>$, so the proof
of the corollary is now completed. \qed 

\vsp
\NI {\bf THEOREM 4.7: } Let $\cla$ be a von-Neumann algebra of type-I with
center completely atomic. Then strong mixing and K-shift properties are 
equivalent. In such a case ( in particular $\cla_0=\clb(\clh)$ )
the following statements
are equivalent:

\NI (a) For any normal state $\psi$, 
$\psi \tau_t \raro \phi_0$ strongly as $t \raro \infty$,

\NI (b) $\clf$ is trivial, 

\NI (c) $\{x \in \cla_0: \tilde{\tau}_t\tau_t(x)=x,\;t \ge 0 \}$ is 
trivial. 

\vsp
\NI {\bf PROOF: } By Proposition 4.6 we only need to show that $\phi_0(J\tau_t(x)J\tau_t(y)) \raro \phi_0(x)\phi_0(y)$ for all $x,y \in \cla$ as $t \raro 
\infty$ whenever it is mixing i.e. $\tau_t(x) \raro \phi_0(x)$ as $t \raro \infty$ in the weak$^*$ topology. 
Since any element can be expressed as linear combination of four non-negative
elements, we assume without loss of generality that $x \ge 0$ and $\phi_0(x)
=1$. For such a choice we note that $\phi_{t}(y) = \phi_0(J\tau_t(x)Jy) \;
y \in \cla$ is a normal state on $\cla_0$ for each $t \ge 0$. By strong mixing
$\phi_t \raro \phi_0$ weakly. Now we use our hypothesis that $\cla_0$ is type-I
with center completely atomic to conclude by a theorem [De] that 
$||\phi_t-\phi_0||_{1} \raro 0$ as $t \raro \infty$. The result follows 
from $|(\phi_t-\phi_0)(\tau_t(y))| \le ||\phi_t-\phi_0||_{1} ||y|| $. 
The last part is now a simple consequence of Theorem 3.10 and Theorem 4.4
and Corollary 3.9. \qed

\vsp
In the proof of Theorem 4.4 we checked that limit$_{t \raro \infty }
\phi_0(JyJ\tilde{\tau}_t(\tau_t(x))) $ exists, in fact the limiting
value for any $x \ge 0$ is less then $||x||\phi_0(JyJ)$ for any $y \ge 0$. Thus
there exists
an element $\cle(x) \in \cla$ so that weak$^*$ limit$_{t \raro \infty}
\tilde{\tau}_t
\tau_t(x) = \cle(x)\; \forall \; x \in \cla$. It is clear that $\cle$ is a
completely positive unital map so that $\phi_0(J\cle(y)Jx)=\phi_0(JyJ\cle(x)$
and $\tilde{\tau}_t\cle\tau_t(x)=\cle$.
However it is not clear whether $\cle^2=\cle$, i.e. a projection in general.
In case $(\tau_t)$ commutes with $(\tilde{\tau}_t)$ then $\cle$ commutes
with $(\tau_t)$ and as well with $(\tilde{\tau}_t)$, thus by taking
limit as $t  \raro \infty $ in second identity we get $\cle^2=\cle$.
Thus in such a case if $\cln=\{x:\tilde{\tau}_t\tau_t(x)=x:\;t \ge 0 \}$ 
we conclude that
$\cle=E$. So we have completed the proof of the following Corollary. 

\vsp
\NI {\bf COROLLARY 4.8: } Let $(\tau_t)$ commutes with $(\tilde{\tau}_t)$. If
$\cln = \{x:\;\tilde{\tau}_t\tau_t(x)=x,\; \; t \ge 0 \}$ then 
$\phi_0(\tau_t(x)\tau_t(y)) \raro \phi_0(xE(y))$ as $t \raro \infty$.

\vsp
We say the system
$(\cla_0,\tau_t,\phi_0)$ is {\it normal } if 
$(\tau_t)$ commutes with $(\tilde{\tau}_t)$ and 
is in {\it detailed balance } if further 
$\tilde{{\cal
L}}(x)-{\cal L}(x)=2i[H,x]$ on a weak$^*$ dense
subalgebra of $\cla_0$, where $H$ is a self-adjoint operator so
that $\alpha_t(x)=e^{itH}x e^{-itH}$ is an automorphism on
$\cla_0$ and ${\cal L}, \tilde{{\cal L}}$ are the generators for $(\tau_t),
(\tilde{\tau}_t)$ respectively. In such a case $(\tau_t)$ commutes 
with $(\alpha_t)$. In the following we investigate results in Theorem 4.1
further. 

\vsp
\NI {\bf THEOREM 4.9: } Let $(\tau_t)$ be in detailed balance with respect
to a faithful normal state $\phi_0$. Then there exists a unique 
unitary
operator $V_0:\clh \raro \tilde{\clh}$ so that 

\NI (a) $V_0:\Omega = \tilde{\Omega}$,

\NI (b) $V_0F_{t]}V^*_0=\tilde{F}_{t]},\;V_0F_{[t}V^*_0=\tilde{F}_{[t}$ for
$t \in \IR$,

\NI (c) $V_0S_tV_0=\tilde{S}_{t} \; t \in \IR $ 

\NI (d) $V_0j_t(\alpha_t(x))V_0^* = \tilde{j}_t(\alpha_{-t}(x))$.

\NI (e) $R_0F_{t]}R^*_0=F_{[-t}$ and $R_0F_{[t}R^*_0 = F_{[-t}$ where $R_0=
U^*_0V_0$.

\vsp
\NI {\bf PROOF : } Since $\tau_t\alpha_t=\tilde{\tau}_t\alpha_{-t}$ for
all $t \ge 0$ and $(\alpha_t)$ commutes with both the semigroup $(\tau_t)$ 
and $(\tilde{\tau}_t)$ 
by (2.2) we check that 
$$V_0:j_{t_1}(\alpha_{t_1}(x_1)) ...j_{t_n}(\alpha_{t_n}(x_n))
\omega=\tilde{j}_{t_1}(\alpha_{-t_1}(x_1)) \tilde{j}_{t_2}..
\tilde{j}_{t_n}(\alpha_{-t_n}(x_n)) \tilde{\Omega}, $$  
is indeed an isometry on total sets generated by the cyclic vectors. Hence 
$V_0$ has a unique extension
to $\clh \raro \tilde{\clh}$. That $V_0$ satisfies (a) -(e) are now routine
work. Uniqueness follows by the cyclic property of the vectors $\Omega$ for 
$\clh$
and $\tilde{\Omega}$ for $\tilde{\clh}$. \qed

\vsp
\newsection{ Quantum mechanical master equation:}

\vsp
We say a normal Markov semigroup $(\tau_t)$ on $\cla_0$ is norm continuous if
limit$_{t \raro 0 }||\tau_t-I||=0$. In such a case the generator $\cll$
is a bounded operator on $\cla$ and can be described [GoKoSu,Lin,CrE] by
\be
\cll(x)= Y^*x +xY + \sum_{k \ge 1} L_k^*xL_k
\ee
where $Y \in \cla_0$ is the generator of a norm continuous
contractive semigroup on $\clh_0$ and $L_k,\;k \ge 1$ is a
family of bounded operators so that $\sum_k L^*_kxL_k \in
\cla_0$ whenever $x \in \cla_0$. However this choice $(Y,L_k,\;k
\ge 1)$ is not unique. Conversely, for any such a family $(Y,L_k)$
with $Y \in \cla_0$ and $L_kxL^*_k \in \cla_0,\;\forall x \in \cla_0$, there
exists a unique Markov semigroup $(\tau_t)$ with $\cll$ as its generator.
There are many methods to show the existence of a Markov 
semigroup $(\tau_t)$ with $\cll$ as it's generator [Da3,MoS,ChF].
Here we describe one such
a method [ChF].
\vsp
We consider the following iterated equation :

\vsp
\begin{eqnarray}
\tau^0_t(x) & = & e^{tY^*}xe^{tY}  \\
\tau^{(n)}_t(x) & = & e^{tY^*}xe^{tY} + \int^t_0e^{(t-s)Y^*}\Phi(
\tau^{(n-1)}_s(x))e^{(t-s)Y}ds
,\;n \ge
1 
\end{eqnarray}
where $\Phi(x)=\sum_k L^*_kxL_k$. It is simple to check for $x \ge 0$ that
$$ 0 \le \tau^{n-1}_t(x) \le \tau^n_t(x) \le ||x||\;I,\;\forall t \ge 0$$
Thus we set for $x \ge 0,\;\tau_t(x)=limit_{n \raro \infty} \tau^{(n)}_t(x)$
in the weak$^*$ topology. For an arbitrary element we extend it by linearity. 
Thus we have 
\be
\tau_t(x) =  e^{tY^*}xe^{tY} + \int^t_0e^{(t-s)Y^*}\Phi(\tau_s(x))e^{(t-s)Y}ds
\ee
for any $x \in \cla_0$.

\vsp
In such a case [Ev], it is simple to check that $\cln$ is trivial if and 
only if $\{x \in \cla_0:[x,H]=0,\;[x,L_k] =0 \forall k \ge 1 \}$ is trivial. 

The following simple but important result due to Fagnola-Rebolledo [FR3].

\vsp
\NI {\bf THEOREM 5.1:} A projection $p$ is sub-normal if and only if 
$(1-p)Yp=0$ and $(1-p)L_kp=0$ for all $1 \le k \le \infty$. 

\vsp
\NI {\bf PROOF:} For a proof and a more general result we 
refer to [FR3]. \qed

\vsp
\NI {\bf THEOREM 5.2:} Let $p$ be a sub-normal projection and $y=\mbox{s.lim}_
{t \raro \infty}\tau_t(p)$. For any $z \in {\cal B}(\clh_0)$ following are 
equivalent:

\NI (a) $yz=0$ 

\NI (b) $pz=0,\;pL_{i_1}L_{i_2}....L_{i_n}z=0$ for all $0 \le i_m \le
\infty$, $1 \le m \le n$ and $n \ge 1$, where $L_0=Y$.

\vsp
\NI {\bf PROOF:} $yz=0$ if and only if $z^*\tau_t(p)z=0$ for all 
$t \ge 0$. Now by (5.2) we have $z^*\tau_t(p)z=0$ if and only if 
$z^*e^{tY^*}pe^{tY}z=0$ and 
$z^*\Phi(\tau_t(p))z=0$ for $t \ge 0$. Thus we have $pe^{tY}z=0$ and also 
$z^*\Phi(\tau_t(p))z=0$ for all $t$. Thus in particular we have 
$z^*\Phi(p)z=0$, hence $pL_kz=0$ for all $k \ge 0$. We go now by induction 
on $n$, we check if $z'=pL_{i_1}L_{i_2}....L_{i_n}z$ then $yz'=z'$ thus (a) 
implies (b). For the converse statement, we check that derivative of any 
order at $t=0$ of $z^*\tau_t(p)z$ vanishes, thus constant which is zero. \qed   

\vsp
Thus the zero operator is the only element $z$ that satisfies (b) 
if and only if the closure of the range of $y$ is the entire Hilbert 
space. Thus $p$ together with $L^*_{i_1}L^*_{i_2}....L^*_{i_n}p$ where 
$0 \le i_m \le \infty$,$1 \le m \le n$ and $n \ge 1$ will generate the 
Hilbert space if and only if $y$ is one to one. In particular we find this 
property is a necessary condition for $y$ to be $1$. In general the 
condition is not a sufficient one. Once more one can construct a counter 
example in birth and death processes where no population is 
an absorbing state and birth and death rates are such that the population will 
extinct with positive probability but need not be $1$. We omit the details. 
However it seems reasonable to ask whether this condition is sufficient for 
finite dimensional Hilbert space ${\cal H}_0$.  

\vsp
Now onwards we assume that $\phi_0$ is faithful. Many important class of 
example [Fr1,FR1,AcM2,Ma,MZ1,MZ2,MZ3] do admit a faithful normal invariant 
state. We check that $\clf_s \subset \{x:\;[L_k,x]=[L_k,x^*]=0,\;k \ge 1 \}$ 
and moreover equality hold if $\{x: \;[L_k,x]=[L_k,x]=0 \}$ is invariant by 
$(\tau_t)$.

\vsp
In case $(\tau_t)$ is only weak$^*$ continuous, the problem in it's complete 
generality is open. For an application to diffusion processes we refer to [Mo].
However a suitable modification of the method
outlined above or a perturbation method can be employed for 
$(Y,L_k)$ unbounded when $\cla_0 = \clb(\clh_0)$. To that end we 
assume [Da 3,ChF,MoS] the following:

\NI (a) $(Y,\;\cld(Y)$ is the generator of a strongly continuous
semigroup with domain $\cld(Y)$;

\NI (b) $L_k$ are closed operator with domains $\cld(L_k) \sbs \cld(Y)$ so
that for $f,g \in \cld(Y)$
$$<f,Yg>+<Yf,g>+\sum_k<L_kf,L_kg>=0;$$

\NI (c) $\IC = \{ x \in \clb(\clh_0):
<f,xYg>+<Yf,xg>+\sum_k<L_kf,xL_kg>=<f,xg> \}$

\vsp
\NI {\bf THEOREM 5.3:} There exists a unique weak$^*$
continuous Markov semigroup $(\tau_t)$ on $\clb(\clh_0)$ with
the generator $\cll$ given by
$<f,\cll(x)g>=<f,xYg>+<Yf,xg>+\sum_k<L_kf,xL_kg>$
for all $f,g \in \cld(Y)$. Moreover the
domain of $\cll$ contains the dense $*$-algebra $\{
(\lambda-Y^*)^{-1}x(\mu-Y)^{-1}: x \in
\clb(\clh_0),\;\lambda,\mu > 0 \}.$  In such a case $\clf
\sbs 
\{ x \in \clb(\clh): [L_k,R_{\lambda}(x)]=0,\;\lambda > 0,\;
1 \le k < \infty \}.$ 

\NI {\bf PROOF:} Let $x \in \clf$. We first check that $\tau_t(x) \in \clf$,
thus $R_{\lambda}(x) \in \clf$. So $R_{\lambda}(x) \in \cld(\cll)$, the domain
of $\cll$ and as well an element in $\clf$. Hence $\cll(R_{\lambda}(x))y+
x\cll(R_{\lambda}(y))=\cll(R_{\lambda}(x)R_{\lambda}(y))$ for
any $x,y \in \clf$. From the explicit relation we find that 
$<f,[R_{\lambda}(x),L_k]^*[R_{\lambda}(x),L_k]g>=0$ for $f,g \in \cld(Y)$ and
$k \ge 1$. In other words $[R_{\lambda}(x),L_k]=0$ has a bounded extension 
and it's value is zero for each $\lambda > 0$.  \qed

\bigskip
{\centerline {\bf REFERENCES}}

\begin{itemize} 

\bigskip
 
\item{[AcM1]} L. Accardi, Anilesh Mohari:
On the structure of classical and quantum  flows
Preprint Volterra, N.167 Febbraio 1994,
Journ. Funct. Anal. 135 (1996) 421--455

\item{[AcM2]} Accardi, L., Mohari, A.: Time reflected 
Markov processes. Infin. Dimens. Anal. Quantum Probab. Relat. Top., vol-2
,no-3, 397-425 (1999).

\item {[Ar]} Arveson, W.: Pure $E_0$-semigroups and absorbing states. 
Comm.Math.Phys 187 , n0.1, 19-43, (1997)

\item {[BhP]} Bhat, R., Parthasarathy, K.R.: Kolmogorov's
existence theorem for Markov processes on $C^*$-algebras, Proc.
Indian Acad. Sci. 104,1994, p-253-262.

\item {[BrR]} Bratelli, O., Robinson, D.W. : Operator algebras
and quantum statistical mechanics, I,II, Springer 1981.

\item {[ChF]} Chebotarev, A.M., Fagnola, F. Sufficient conditions for 
conservativity of minimal quantum dynamical semigroups, J. Funct. Anal. 
153 (1998) no-2, 382-404.

\item{[CrE]} Christensen, E., Evans, D. E.: Cohomology of
operator algebras and quantum dynamical semigroups, J.Lon. Maths
Soc. 20(1970) 358-368.

\item{[Ci]} Cipriani, F.: Dirichlet form and Markovian semigroups on
standard forms of von Neumann algebras, J. funct. Anal. 147 (1997) n0-2, 259-300.

\item{[Da1]} Davies, E.B.: One parameter semigroups, Academic
Press, 1980.

\item{[Da2]} Davies, E.B.: Quantum theory of open systems,
Academic press, 1976.

\item{ [Da3}] Davies, E.B.: Quantum dynamical semigroups and the neutron diffusion equation. Rep. Math. Phys. 11 (1977), no-2, 169-188. 

\item{[De]} Dell'Antonio, G.F. : On the limit of sequences of normal states,
Comm. Pure Appl Math 20 (1967) 413-429. 

\item{[Ev]} Evans, D.E.: Irreducible quantum dynamical
semigroups, Commun. Math. Phys. 54, 293-297 (1977).

\item{[EL]} Evans, D.E., Lewis, J.T.: Dilations of irreversible
evolution in algebraic quantum theory, Dublin institute of
advance studies, (1977).

\item{[FR1]} F.~Fagnola, R.~Rebolledo.
The approach to equilibrium of a class of quantum dynamical semigroups
Inf. Dim. Anal. Q. Prob. and Rel. Topics, 1(4):1--12, 1998
 
\item {[FR2]} F.~Fagnola and R.~Rebolledo.
On the existence of invariant states for quantum dynamical semigroups
J.Math.Phys., 42, 296-1308, 2001.
 
\item {[FR3]} F.~Fagnola, R.~Rebolledo.
Subharmonic projections for a Quantum Markov Semigroup.
Preprint PUC/FM-04/2000, Santiago.
 
\item{[MZ1]} Majewski A.W. and Zegarlinski B., {\it On quantum stochastic
dynamics }, Markov Proc. and Rel. Fields {\bf 2} (1996) 87--116
 
\item{[MZ2]} Majewski A.W. and Zegarlinski B., {\it Quantum Stochastic
Dynamics II}, Rev. Math. Phys. {\bf 8} (1996) 689--713
 
\item{[MOZ1]} A. Majewski, Olkiewicz R. and Zegarlinski B., {\it %
Dissipative Dynamics for Quantum Spin Systems on a Lattice}, J. Phys. A.:
Math. Gen. {\bf 31} (1998) 2045--2056
 
\item{[MOZ2]}  A. Majewski, Olkiewicz R. and Zegarlinski B., {\it %
Construction and Ergodicity of Dissipative Dynamics for Quantum Spin Systems
on a Lattice}, pp.112 - 126 in ''Frontiers in Quantum Physics'' Eds. S.C.
Lim, R. Abd-Shukor, K.H. Kwek , Springer-Verlag 1998
 
\item{[MOZ3]}  A. Majewski, Olkiewicz R. and Zegarlinski B., {\it %]
Stochastic Dynamics of Quantum Spin Systems}, pp. 285--295 {\it Banach
Center Publications}, Vol. {\bf 43}, 1998, Quantum Probability 97, Eds R.
Alicki, M. Bozejko and W. A. Majewski

\item{[FKGV]} A. Frigerio, A. Kossakowski, V. Gorini, M. Verri:
"Quantum detailed balance and KMS condition."
Commun. Math. Phys. 57 (1977) 97-110.
Erratum: Commun. Math. Phys. 60 (1978) 96.
 
\item{[Fr1]} Frigerio, A.: Stationary states of quantum dynamical
semigroups. Commun. Math. Phys. 63, 269-276 (1978).

\item{[Fr2]} Frigerio, A., Gorini, V. : Markov dilations and quantum detailed
balance. Commun. Math. Phys. 93 , 517-532 (1984).  

\item{[GoKoSu]} Gorini, V., Kossakowski, A., Sudarshan, E.C.G. : Completely
positive dynamical semigroups of n-level systems, J. Math. Phys. 17,
821-825 (1976).

\item{[Lin]} Lindblad, G. : On the generators of quantum
dynamical semigroups, Commun.  Math. Phys. 48, 119-130 (1976).

\item{[Mo]} Mohari, A.: Ergodicity of Homogeneous Brownian flows. Preprint.

\item{[MoS]} Mohari, A., Sinha, K.B.: Stochastic dilation of minimal quantum dynamical
semigroup. Proc. Indian Acad. Sci. Math Sci. vol-102 (1992), no-2, 159-173. 

\item{[OhP]} Ohya, M., Petz, D.: Quantum entropy and its use, Text
and monograph in physics, Springer-Verlag.

\item{[Par]} Parry, W. :  Topics in ergodic theory, Cambridge
University press, 1981.

\end{itemize}

\end{document}